\documentclass[final,5p,times]{elsarticle}
\usepackage{bm} 
\usepackage{color}
\usepackage{graphicx}
\usepackage{epsfig}
\usepackage{amssymb}
\usepackage{amsmath} 
\usepackage{amsthm}
\usepackage{amstext}
\usepackage{dcolumn} 
\usepackage{makeidx}
\usepackage{subfigure}
\usepackage{xcolor}

\definecolor{darkgreen}{rgb}{0,.5,0}
\usepackage[colorlinks,urlcolor=blue,filecolor=blue,linkcolor=blue,citecolor=blue]{hyperref}
\journal{Physics Letters A}
\begin{document}
\begin{frontmatter}
\title{Nonlinear optomechanical systems with quasi-periodic and chaotic dynamics}
\author[inst1]{A.P. Saiko\corref{cor1}}
\cortext[cor1]{Corresponding author.}
\ead{saiko@physics.by}
\author[inst1]{G.A. Rusetsky}
\author[inst1]{S.A. Markevich}
\author[inst2]{R. Fedaruk}
\affiliation[inst1]{organization={Scientific-Practical Material Research Centre, Belarus National Academy of Sciences}, addressline={19 P.Brovka str.}, city={Minsk}, postcode={220072}, country={Belarus}}
\affiliation[inst2]{organization={Institute of Molecular Physics, Polish Academy of Sciences}, addressline={17 Smoluchowski str.}, city={Poznan}, postcode={60-179}, country={Poland}}

\begin{abstract}
Model optomechanical systems with  photon-vibration interactions linear, quadratic, and cubic in mechanical displacements are studied under conditions for adiabatic elimination of the photon field. The opportunity of transformation of effective potential describing the dynamics of the mechanical resonator from single-well to double-well is demonstrated. The dynamics of the mechanical resonator is considered in the presence of (i) only linear interaction, (ii) only quadratic interaction, (iii) both linear and quadratic interactions, and (iv) all three interactions, while other parameters of the optomechanical system, the modulation optical field, and the initial conditions remain fixed. Quasiperiodic oscillations of the mechanical resonator in the case (i) are replaced by chaotic ones when the cases (ii) or (iii) are realized. It is interesting that in the presence of all three interactions, the chaotic behavior of the mechanical oscillator becomes quasi-periodic. However, increasing the power of the modulation field again leads to chaos.
\end{abstract}
\end{frontmatter}

\section{Introduction}

The interaction between optical and mechanical modes in optomechanical systems \cite{pp1} results in such effects as optomechanically induced transparency \cite{pp2, pp3, pp4, pp5, pp6, pp7,pp8, pp9}, phonon lasers \cite{pp10, pp11, pp12, Li2020, Sheng2020, Piergentili2021, Grudinin2010, Jing2014, Xie2020}, mechanical resonator (oscillator) cooling \cite{pp13, pp14, pp15, pp16}, photon blockade \cite{pp17, pp19, pp20, pp21}, normal-mode splitting \cite{pp22, pp23}, nonclassical state preparation \cite{pp24, pp25, pp26, pp27}, chaotically spiking attractors \cite{Marino2011}, and dynamical multistability induced by radiation pressure \cite{ppMarquardt}. Introducing the two-laser driving, controllable enhancement of the single-photon optomechanical coupling in a prototypical Fabry-Perot cavity was realized \cite{pp29}. Optomechanical systems have potential applications in quantum information processing \cite{pp30}. The collapse and revival of mechanical and optical oscillations have been considered \cite{pp31, pp32}. The influence of nonlinearities of mechanical oscillators was studied in \cite{pp33, pp34}. The generation of higher order sidebands \cite{pp35}, optomechanical entanglement \cite{pp36}, mechanical quantum squeezing \cite{pp37}, optomechanically induced transparency \cite{pp38, pp39}, normal-mode splitting \cite{pp40}, Kerr and cross-Kerr nonlinearities \cite{pp41} were described.

The optomechanical interaction linear in mechanical displacement ($\sim x$) is usually considered. Since the optomechanical interaction is inherently nonlinear, higher order interactions should be taken into account. For example, quadratic interactions ($\sim x^{2}$) were usually implemented in optomechanical systems using membrane-in-the-middle configurations \cite{pp6, pp42}. Recently, the microwave optomechanical system in the self-oscillating regime was studied taking into account nonlinear interactions up to third order ($\sim x^{3}$) \cite{pp45}. In Ref. \cite{Gao2015}, self-sustained oscillations and limit cycles were studied in the large-displacement regime when the frequency ${\omega _c}(x)$  of the cavity optical mode was not expanded in powers of displacement of the mechanical resonator. Such nonlinearities also result in cross-Kerr effects \cite{pp44} and the collapse and revival of the optical field in optomechanical systems \cite{pp32}.

In recent years, a number of significant studies have been devoted to understanding the transition from classical nonlinear dynamics to chaotic dynamics in optomechanical systems (see, for example, \cite{pp42, pp43}). The obtained results lay the foundation for the application of mechanical micro- and nanoresonators in sensing \cite{ppn5, ppn6}, for generating random numbers and data encryption based on optomechanics, optomechanical logic and chaos computing (see \cite{pp43} and references therein).

Our aim is to study the dynamics of a mechanical oscillator in optomechanical systems with the photon-vibration interactions linear, quadratic, and cubic in mechanical displacements. We focus on the unusual dynamics of the mechanical resonator when the type and magnitude of nonlinearity of the optomechanical interaction change. In particular, our studies show that the simultaneous presence of linear and quadratic optomechanical interactions leads to a chaotic regime of the dynamics of the mechanical resonator (at a fixed choice of the parameters of the optomechanical system and initial conditions). If, for example, due to symmetry restrictions, only the quadratic interaction is preserved, then the behavior of the mechanical resonator is intermediate between quasi-periodic and chaotic. It is interesting that the complex type of optomechanical interaction, when linear, quadratic and cubic couplings are taken into account simultaneously, does not lead to chaos in mechanical oscillations, but only to their multi-quasiperiodicity. Throughout we consider the situation where the cavity decay rate is much larger than the resonant frequency of the mechanical oscillator, and therefore adiabatic elimination of the cavity field is valid.

\section{Theory}

We study a model optomechanical system that consists of a mechanical resonator (oscillator) with mass $m$ and frequency $\omega _{m}$. This oscillator interacts nonlinearly with an optical cavity mode and this interaction includes terms linear, quadratic and cubic in mechanical displacements. The Hamiltonian \cite{pp45} of this system is

\begin{equation}
\label{EQ 1}
H=H_{0} +V+V_{d},
\end{equation}
\[H_{0} =\omega _{c} \hat{a}^{\dag } \hat{a}+\frac{1}{2} (\frac{\hat{p}^{2} }{m} +m\omega _{m}^{2} \hat{x}^{2} ),\]
\[V=-\hat{a}^{\dag } \hat{a}\sum_{n=1,2,3} g_{n} \hat{x}^{n},\]
\[V_{d}^{} =i\varepsilon (\hat{a}^{\dag } e^{-i\omega _{d} t} -H.c.),\]
where $\hat{a}$ ($\hat{a}^{\dag }$) denotes the bosonic annihilation (creation) operators for the cavity mode with $[\hat{a},\hat{a}^{\dag } ]=1$, $[\hat{x},\hat{p}]=i$ (Planck's constant $\hbar =1$), $\hat{x}$ and $\hat{p}$ are the position and momentum operators for the mechanical resonator, $\omega _{c}$ is the cavity frequency, $g_{n}$ is the optomechanical coupling strength, \textit{n} = 1, 2, 3 for linear, quadratic and cubic interactions, respectively. Here $V_{d}$ is the optical driving term with the amplitude $\varepsilon$ and frequency $\omega _{d}$. The three types of coupling presented in the Hamiltonian \eqref{EQ 1} can be realized in an optomechanical system consisting of a drumhead nano-electro-mechanical resonator coupled to a microwave cavity \cite{pp45}.

In the quantum Langevin equations, obtained on the bases of the Heisenberg equations, which are derived from the Hamiltonian \eqref{EQ 1}, we will  make the transition to classical analogies of operators and neglect noise sources \cite{pp455, pp28}. We assume that the decay rate $\kappa$ of the optical field is much larger than the resonant frequency of the mechanical oscillator. Then the optical field variables may be adiabatically eliminated \cite{pp28} from these equations. Finally, on time scales longer than $\kappa ^{-1}$, we obtain the following classical equation for the momentum $p$:
\[\frac{d}{dt} p=-m\omega _{m}^{2} x+\]
\begin{equation} \label{EQ 2}
\frac{\varepsilon ^{2} }{\left[\Delta -\sum_{n=1,2,3} g_{n} x^{n}  \right]^{2} +\kappa ^{2} /4} \frac{d}{dx} \sum_{n=1,2,3} g_{n} x^{n}  -\frac{\gamma }{2} p,
\end{equation}
where $\Delta =\omega _{c} -\omega _{d} $, and $\gamma $ is the phenomenologically introduced mechanical damping rate. Note that at  $\kappa \approx \omega_m$, the adiabaticity condition is violated. In this case the dynamical multistability in high-finesse micromechanical optical cavities has been discovered \cite{ppMarquardt}. Beyond the adiabatic approximation, the effects of radiation pressure have also been studied in \cite{Rokhsari2005, Carmon2005, Carmon2007, Miri2018, Christou2021, gu2024, Zhang2024}. Neglecting the mechanical damping, Eq. \eqref{EQ 2} can be represented in the form $dp/dt=-\partial U_{eff} /\partial x$, where

\begin{equation} \label{EQ 3}
U_{eff} (x)=\frac{m\omega _{m}^{2} }{2} x^{2} +\frac{2\varepsilon ^{2} }{\kappa } \arctan \left[\frac{\Delta -\sum _{n=1,2,3}g_{n} x^{n}  }{\kappa /2} \right].
\end{equation}

So, the dynamics of the mechanical resonator is described by the effective potential $U_{eff} (x)$, in which an additional term is added to the initial harmonic potential. This   term depends on the amplitude and frequency of the driving field, the coupling strengths and the cavity decay rate. Since the presence of damping of mechanical oscillations did not qualitatively affect noticeably the transition of the optomechanical system to quasi-periodic and chaotic motion, we do not take into account in further consideration. This allows us to use the derived effective potential for the mechanical resonator in interpreting the results.

\section{Results and discussion}

\begin{figure}[h]
\centering
\includegraphics[width=8 cm]{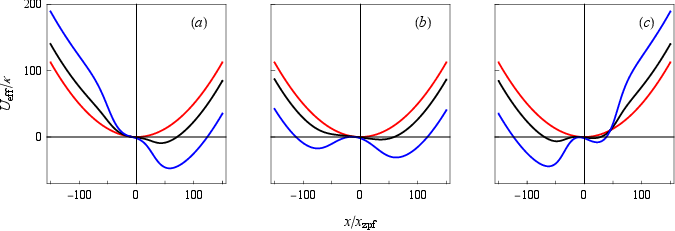}
\caption{The effective mechanical potential at different cubic coupling strengths $g_{3} $ and amplitudes $\varepsilon /\kappa $ of the driving field. The normalized (with respect to $\omega _{m}$) parameters are $\kappa =50.0$, $\omega _{m} =1$, $\Delta =-1.0$, $g_{1} =0.15$ and $g_{2} =0.0075$. Red, black and blue lines were obtained at $\varepsilon /\kappa =0,3$ and $5$, respectively. (a) $g_{3} =0.00025$. (b) $g_{3} =0$. (c) $g_{3} =- 0.00025$.}
\label{fig:Fig1}
\end{figure}

Fig. \ref{fig:Fig1} shows effects of the value and sign of the cubic optomechanical coupling strength $g_{3}$ and the amplitude $\varepsilon$ of the optical driving field on the shape of the effective mechanical potential $U_{eff}(x)$ (Eq. \eqref{EQ 3}). To express all constants in energy (frequency) units, it is necessary to make the following substitution: $g_{i} \to g_{i} (1/2m\omega _{m} )^{i/2} $, where $(1/2m\omega _{m} )^{1/2} \equiv x_{zpf} $ is the size of mechanical zero-point fluctuations. At the positive values of $g_{3} $ (Fig. \ref{fig:Fig1}\emph{a}), the effective potential has one minimum, and the mechanical oscillator has one stable state of equilibrium. With a increase in $\varepsilon$, the minimum shifts towards higher values of $x$. When the cubic coupling strength $g_{3}$ becomes negative (Fig. \ref{fig:Fig1}\emph{c}), the effective mechanical potential is transformed into a double-well one, and the mechanical oscillator can be in a metastable state. At $g_3 = 0$ and $g_1 =0$ the potential has a symmetric two-minimum shape for arbitrary values of $\varepsilon$, but at $g_1 \neq 0$, an asymmetry appears in the location of the minima (Fig. \ref{fig:Fig1}\emph{b}). This asymmetry increases with increasing the linear optomechanical coupling strength and the amplitude of the optical driving field. In all three cases shown in Fig. \ref{fig:Fig1} ($g_3$ is positive, zero and negative), in the absence of a driving field ($\varepsilon = 0$) the potential is, of course, harmonic, i.e. equal to $m\omega _{m}^{2} {x}^{2} /2$. Note that asymmetric double-well potentials similar to those discussed here are often encountered in condensed matter physics problems \cite{pp46}.

As in \cite{pp28}, we assume that the input power of the driving field is modulated with frequency $\Omega$, and $\Omega \ll \kappa $. Therefore, one can consider the non-stationary behavior of the mechanical resonator with a time-dependent potential if in Eq. \eqref{EQ 2} the input power parameter $\omega _{d} \varepsilon ^{2} /2\kappa $ is changed as:

\begin{equation} \label{EQ_7}
\frac{\omega _{d} }{2\kappa } \varepsilon ^{2} \to \frac{\omega _{d} }{2\kappa } (\varepsilon ^{2} -\varepsilon _{M}^{2} \sin \Omega t),\, \, \varepsilon _{M} \le \varepsilon ,
\end{equation}
where $\varepsilon _{M} $ is the modulation field amplitude.

When carrying out numerical calculations using Eqs. \eqref{EQ 2} and  \eqref{EQ_7} to construct phase portraits, the time was maintained from 0 to 0.01 s. The Poincaré section is the position of the particle in the phase space ($\dot x, x$ ) at times ${t_j} = (2\pi /\Omega )j$, where integer values   varied from 1 to 1800 over 0.1  s. In phase space, the evolution of a dynamic oscillatory system is described by a phase trajectory. The set of initial states of this system (the set of representative points in phase space) in the process of evolution corresponds to a set of phase trajectories. If the oscillatory process is established, then all the representative points of the phase plane are collected on an attractor in the form of a closed curve, which is called a limit cycle. Such a limit cycle on the Poincaré section corresponds to a point. The  $m$-cycles correspond to $m$  points on the section, i.e. the oscillatory regime becomes quasi-periodic. The transition to a chaotic set of points on the Poincaré section indicates a transition to an oscillatory regime called dynamic chaos.

\begin{figure}[h]
\centering
\includegraphics[width=7.88 cm]{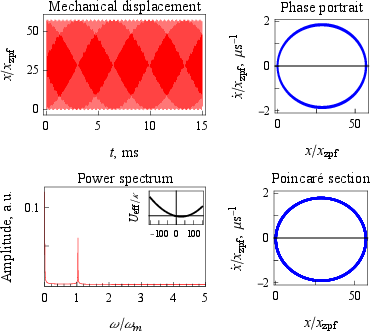} \caption{Dynamics of the mechanical oscillator when the optomechanical interaction contains only a linear term ($g_{1} =0.15$, $g_{2} =g_{3} =0$). The inset shows the effective mechanical potential. The normalized (with respect to $\omega _{m}$) parameters are  $\kappa =50.0$, $\Delta = -1.0$, $\varepsilon /\kappa = 5$, $\varepsilon _{M} /\varepsilon =0.20007$, and  $\Omega=1.8$.}
\label{fig:Fig2} \end{figure}

The dynamics of the mechanical resonator in the optomechanical system when the Hamiltonian of the optomechanical interaction contains only the term that is linear in mechanical displacements is illustrated in Fig. \ref{fig:Fig2}. In this case, the effective potential of the mechanical resonator has a minimum that is somewhat shifted toward the positive region. The resonator oscillations are of a quasi-periodic nature. Their power spectrum contains a high-intensity main peak at the frequency $\omega _{m} $ and a significantly weaker one at the driving field frequency $\Omega $, as well as a barely noticeable second harmonic of $\omega _{m} $ . The Poincar\'{e} section has the shape of an almost ideal circle, shifted, like the phase portrait, into the positive region.

\begin{figure}[h]
\centering
\includegraphics[width=7.88 cm]{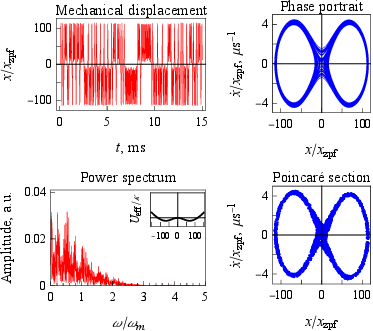} \caption{Dynamics of the mechanical oscillator when the optomechanical interaction contains only quadratic term ($g_{1} =g_{3} =0$, $g_{2} =0.0075$). The inset shows the effective mechanical potential. Other parameters are the same as in Fig. \ref{fig:Fig2}.}
\label{fig:Fig3}
\end{figure}

If the optomechanical interaction is quadratic in mechanical displacements, the corresponding anharmonic effective potential of the mechanical resonator has a symmetric double-well shape centered on $x = 0$. The shape of the power spectrum, phase portrait and Poincar\'{e} section indicate the emerging transition from dynamic to chaotic behavior (Fig. \ref{fig:Fig3}). The time dependence of mechanical displacement shows that at first the mechanical resonator oscillates predominantly in the right well with infrequent transitions to the left well. Then, over time, the situation changes to the opposite, and a similar alternation of changes in the oscillatory positions of the mechanical system continues in the future. A structured continuous power spectrum is formed, and the phase portrait and Poincaré section take on the appearance of a slightly blurred figure eight, elongated along the $x$ axis.

\begin{figure}[h]
\centering
\includegraphics[width=7.88 cm]{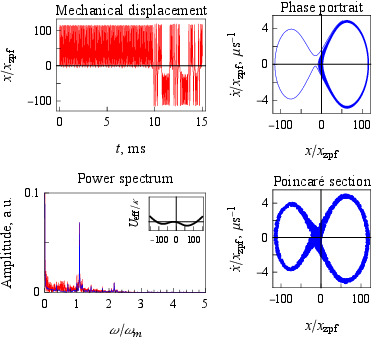} \caption{Dynamics of the mechanical oscillator when the optomechanical interaction contains linear and quadratic terms ($g_{1} =0.15$, $g_{2} =0.0075$ and $g_{3} =0$). The blue line in the power spectrum is the Fourier spectrum of mechanical displacement taken over the interval from 0 to 10 ms. The inset shows the effective mechanical potential. Other parameters are the same as in Fig. \ref{fig:Fig2}.}
\label{fig:Fig4}
\end{figure}

When only linear and quadratic terms in mechanical displacements are in the optomechanical interaction, the symmetry of the double-well effective potential is broken: the right well drops slightly lower than the left one. The resonator oscillations remain chaotic in nature, and, compared to the previous case, the phase portrait and Poincaré section have the appearance of a slightly modified figure eight (Fig. \ref{fig:Fig4}). The time dependence of mechanical displacement indicates that, during the first ten milliseconds, the mechanical resonator performs quasi-harmonic oscillations in the right well. This is confirmed by the corresponding part of the power spectrum shown by the blue line. Next, there is an alternation of the presence of the chaotically oscillating mechanical resonator in the left or right wells with its somewhat predominant presence in the lower right well. The full power spectrum takes on a structured continuous shape.

\begin{figure}[h]
\centering
\includegraphics[width=7.88 cm]{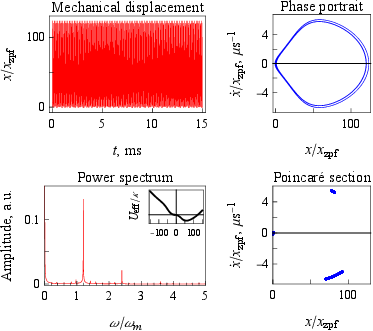} \caption{Dynamics of the mechanical oscillator. $g_{1} =0.15$, $g_{2} =0.0075$ and $g_{3} =0.00025$. The inset shows the effective mechanical potential. Other parameters are the same as in Fig. \ref{fig:Fig2}.}
\label{fig:Fig5}
\end{figure}

If the optomechanical interaction contains linear, quadratic and cubic terms in mechanical displacements, the effective potential of the mechanical resonator for the selected set of system parameters has a global minimum and an inflection point on its relief (Fig. \ref{fig:Fig5}). In the power spectrum, in addition to the intense main peak at frequency $\omega _{m} $ and the weaker one at frequency $\Omega $, there are their harmonics and subharmonics, and all these peaks have their satellites on the left and right. The phase portrait consists of two fairly narrow closed bands inserted into each other in the positive region and connected near a point ($\dot{x}, x$)=(0,0). On the Poincar\'{e} plane, this phase portrait corresponds to three different lines, one of which is located near the point ($\dot{x}, x$)=(0,0), the second, longer one, is in the upper part of the Poincaré plane, and the third, the longest one, is in the lower part. Thus, a 3-quasiperiodic motion of the mechanical resonator is realized (Fig. \ref{fig:Fig5}).

\begin{figure}[h]
\centering
\includegraphics[width=6.86 cm]{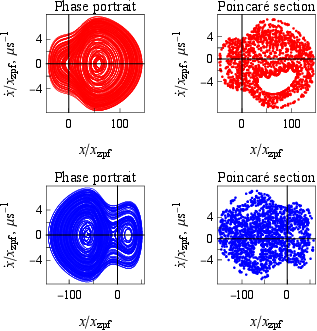} \caption{Phase portraits and Poincar$\acute{e}$ sections for the mechanical resonator at $\varepsilon _{M} /\varepsilon =0.5001$, $g_{1} =0.15$, $g_{2} =0.0075$, $g_{3} =0.00025$ (red plots) and $g_{3} =- 0.00025$ (blue plots). Other parameters are the same as in Fig. \ref{fig:Fig2}.}
\label{fig:Fig6}
\end{figure}

It is interesting that the presence of all three types of the optomechanical interaction does not lead to chaotization of the oscillations of the mechanical resonator, as it happens in the presence of both linear and quadratic interactions (Fig. \ref{fig:Fig4}), or only quadratic interaction (Fig. \ref{fig:Fig3}). When adding the cubic optomechanical interaction to the linear and quadratic terms, the right global minimum of the effective potential for the mechanical oscillator becomes deeper and flatter. As a result, at a given low modulation (${\varepsilon _M}/\varepsilon  = 0.20007$), the mechanical oscillations are carried out within the limits of only this global minimum with the realized quasi-periodic motion and hints of a transition to chaotic behavior. A relatively small increase the modulation (${\varepsilon _M}/\varepsilon  = 0.5001$) takes the mechanical oscillator out of the global minimum and makes it oscillate in a two-minimum potential, leading to chaotization of its movement (see Fig. \ref{fig:Fig6}).

The transition to chaotic behavior of the mechanical resonator significantly depends on the amplitude of the modulating optical field. Fig. \ref{fig:Fig6} shows phase portraits and Poincaré sections for the mechanical resonator when the amplitude of this field was increased by a factor of 2.5 compared to the case shown in Fig. \ref{fig:Fig5}. The results are given for the optomechanical system with $g_3 > 0$ and $g_3 < 0$. At $g_3 > 0$, the minimum of the effective potential is shifted to the positive region of mechanical displacements (see Fig.  \ref{fig:Fig1}\emph{a}), and, accordingly, the phase portrait and the Poincaré section illustrate the chaotic motion of the mechanical resonator mainly in this region. When $g_3 < 0$ and the two-minimum effective potential with the left global minimum is realized (see Fig. \ref{fig:Fig1}\emph{c}), the phase portrait and the Poincaré section are mainly in the negative region of mechanical displacements, i.e. chaotic movements occur predominantly in the left global minimum.

\section{Conclusion }

We have studied model optomechanical systems with the photon-vibration interactions linear, quadratic, and cubic in mechanical displacements. It was found that under the conditions of adiabatic elimination of the optical field, the dynamics of the mechanical oscillator is determined by some effective potential. The transformation of this potential from single-well to double-well and bistable behavior of the mechanical resonator were demonstrated. The transition of the mechanical resonator from dynamic to chaotic behavior was shown when ``switching on'' linear, quadratic and cubic (or their combination) optomechanical interactions in mechanical displacements with a certain choice of other parameters of the system under study and initial conditions. Maintaining this set of parameters and initial conditions, we have found that the transition to the chaotic oscillation regime of the mechanical resonator is realized in the presence of both linear and quadratic interactions. In this case, the presence of only quadratic interaction is a boundary factor for the transition from dynamic to chaotic behavior. In any case, the transition to chaotic behavior is not realized without the presence of a quadratic term in mechanical displacements. The found features of nonlinear dynamics of optomechanical systems are significant for understanding physical processes in open quantum systems as well as for possible practical applications, including sensors based on mechanical micro- and nanoresonators \cite{ppn5, ppn6}, optomechanical logic and chaos computing (see \cite{pp43} and references therein)

\end{document}